# Magnetization ground states and phase diagrams for a nano-sized Co hollow sphere : an onion-type magnetization state

Desheng Kong, Siming Wang, and Chinping Chen[a]

Department of Physics, Peking University, Beijing 100871, People's Republic of China

Abstract

The magnetization ground states (MGSs) for a nanosized Co hollow sphere, with the outer radius, $R < 50$ nm, have been studied numerically by micromagnetic simulation using object oriented micromagnetic framework (OOMMF). In addition to the originally known single domain and vortex-curling states, a three dimensional "onion" state with a corresponding analytical expression is proposed and confirmed as one of the ground states. Two phase diagrams, one for a single crystalline and the other for a polycrystalline nanosphere, are obtained for the three MGSs. The result reveals that the magnetic anisotropy has a significant effect on the phase line in the diagrams. The finite temperature effect and the blocking properties of the nanosphere for the magnetization reversal are discussed.

## 1. Introduction

With the progresses on the controlled synthesis of nanoelements in material science, it is possible to synthesize nanoparticles with various designed shapes and structures. In particular, many works have demonstrated the methods and mechanisms of synthesizing nanosized hollowspheres with diameters ranging from tens to hundreds

---

[a] Corresponding author. Electronic-mail : cpchen@pku.edu.cn, Phone : +86-10-62751751

of nanometers.[1-4] Recently, nanosized ferromagnetic hollow spheres,[5-7] chains of submicron-sized Co hollow spheres,[8,9] and chains of Co hollow spheres with each cavity void containing a solid nanoparticle[10] were successfully produced by experiments. It is therefore possible to tailor the magnetic properties of nanoparticles by varying the shell thickness of the hollow structure in addition to the particle size, thus, breeding basic research interests with great application potentials.

The configuration of the magnetization ground states (MGSs) is one of the most important properties for the three dimensional (3D) magnetic hollow sphere. Goll *et al*.[11] have performed an analytical study on the MGSs for a nanosized hollow sphere. By comparing the total energy of the four well-defined magnetization states, *i.e.*, the single domain (SD), two domain, four domain and vortex-curling (VC) states, they have reached the conclusion that the SD and the VC states are the MGSs for a nanosized hollow sphere of Co with the outer radius, $R < 50$ nm. These are simple magnetization states without complicated domain wall structures since the formation of domain wall takes more exchange energy than the corresponding reduction in the magnetostatic energy with the particle size in the nanoscale region under consideration.

Similar studies on the SD and VC states for the two dimensional (2D) counterpart, *i.e.*, the nanodisk or nanoring, have drawn much attention as well.[12-15] Besides the SD and VC states, a third candidate, the so-called "onion" state (O-state), has been proposed for a 2D nanoring, and confirmed by micromagnetic simulations and experiments.[16-20] These works show that for a 2D O-state, the magnetization tends to follow the curvature in each half of the nanoring with opposite sense of rotation and thus forms a two domain magnetization structure. Recently, Landeros *et al*[21] proposed an analytical formula in a theoretical work to describe the O-state of a 2D nanoring. Consequently, a phase diagram is obtained for the O, SD, and VC states.

In the present investigation, we have introduced a 3D O-state as a MGS for a nanosized hollow sphere in addition to the SD and VC states. The magnetization configuration of a 3D O-state is a variation from the SD state with the direction of magnetization at each local point deviating away from the homogeneously parallel distribution. A criterion is then obtained by extending the analytical expression originally proposed for a 2D nanoring[21] to discern the subtle difference between an O-state and a SD state for a 3D nanosphere. By calculating the energy for the magnetization configuration of the SD, VC and O-states, two phase diagrams are obtained. One is for a single crystalline Co nanosphere with the magnetocrystalline anisotropy, $K_{mag}$ = 4.0×10$^5$ J/m$^3$ and the other, for a polycrystalline nanosphere, assuming $K_{mag}$ = 0. The phase diagrams cover the result with the inner radius, $R_{in}$, ranging from zero, *i.e.*, $\Lambda_R = R_{in}/R = 0$ for a solid particle, to a very thin-shelled hollow sphere, *i.e.*, $\Lambda_R = 0.8$.

## 2. Calculation and energy minimization procedure

The numerical calculation has been performed based on the object oriented micromagnetic framework (OOMMF) package from NIST.[22] The method of finite difference calculation adopted by this program has been demonstrated to be effective for investigating the magnetic properties of nanomagnets.[12,13,17-19] For the modeling, we select the parameters of bulk hcp Co, *i.e.*, the saturation magnetization, $M_S$ = 1.4×10$^6$ A/m (~157.3 emu/g) and the exchange stiffness constant, $A$ = 30 pJ/m.[11] In addition, for the single crystal hollow sphere, the magnetocrystalline anisotropy of bulk hcp Co is used, $K_{mag}$ = 4.0×10$^5$ J/m$^3$,[11] whereas for the polycrystalline, $K_{mag}$ = 0.[19,23] By taking the computing capacity into account, different lattice cell size has been used in the calculation depending on the size of hollow spheres, which is, 1 × 1 × 1 nm$^3$ for the spheres with $R$ > 30 nm, 0.5 × 0.5 × 0.5 nm$^3$ for 10 nm < $R$ < 30 nm, and 0.25 × 0.25 × 0.25 nm$^3$ for $R$ < 10 nm. This is valid in that all of the cell sizes are far smaller than the exchange length of Co, $L_{ex}^{Co} = \sqrt{2A/\mu_0 M_S^2}$ ~5 nm.[13] Although the overall computational scheme is robust, a numerical error is unavoidably introduced by the effect of surface roughness arising from the discrete nature of the cubic lattice cells, as discussed in Refs 12 and 13. It becomes more significant as the shell thickness reduces to the value comparable to the lattice cell size. According to our calculation, the error is expected to be small since the variation in total energy calculated by further reducing the cell size is usually smaller by three orders of magnitude than the calculated total energy itself. In a few cases, the difference is about 1%.

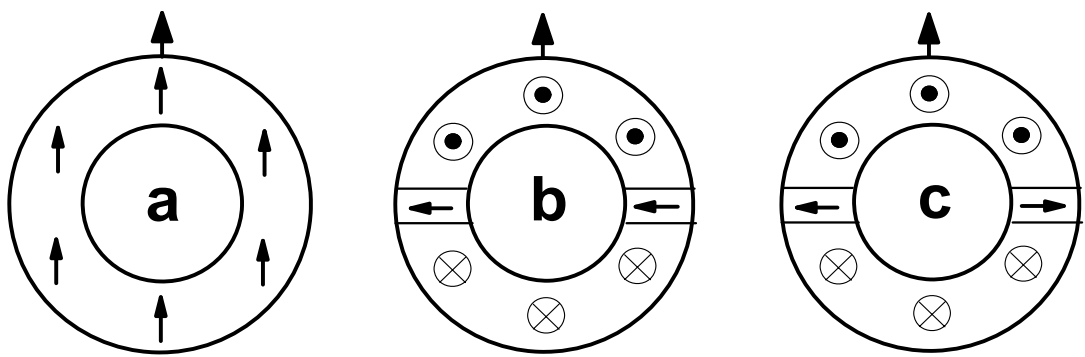

FIG. 1 Simple diagrams for a) the SD state, b) the PVC state, and c) the AVC state. The arrow on top of the hollow sphere pointing upward indicates the direction of easy axis for $K_{mag} > 0$. The magnetization direction in (a) is represented by the arrows inside the shell. In (b) and (c), the magnetization vector is pointed out of the paper in the upper half of the sphere, and into the paper, in the lower half. The two arrows in the two vortex cores in both (b) and (c) are for the local direction of magnetization.

For the energy minimization calculation to determine the MGS, the SD or VC state is selected as the initial magnetization configuration to start with. For the SD state, as shown by the simple diagram in Fig. 1(a), the direction of magnetization, as shown by the arrows in the shell, is parallel to the easy axis of the single crystal nanosphere of Co with $K_{mag} > 0$, whereas it points in any arbitrary direction for the polycrystalline nanosphere with $K_{mag} = 0$. The direction of easy axis is pointed upward, as represented by the arrow on top of the hollow sphere. On the other hand, for the VC state as shown in Fig. 1(b) or 1(c), the magnetization forms a flux closure configuration. The direction of magnetization in the upper half of the sphere is rotating out of the paper plane, and the lower half, into the paper plane. In the cap regions near the two poles, vortex cores appear. There are two possible orientation arrangements for this pair of magnetization vortex cores. One is in parallel, termed as the parallel VC (PVC) state, as shown by the two arrows in Fig. 1(b). The other is in antiparallel arrangement, termed as the antiparallel VC (AVC) state, as shown by the two arrows in Fig. 1(c). In the case of $K_{mag} > 0$, it has been pointed out explicitly by a previous study that the total energy is higher with the orientation of the vortex plane perpendicular to the easy axis than in parallel,[11] *i.e.*, the lower energy state corresponds to the one with the axial direction of the two vortex cores perpendicular to the easy axis. In the case of $K_{mag} = 0$, the easy axis does not exist. Thus, there is no preferential orientation of the vortex cores for the total energy. In addition, it has been shown that the magnetic coupling energy between these two cores is lower for the PVC state than the AVC state.[24] To determine the MGS, an energy minimization process based on the conjugate gradient algorithm embedded in the OOMMF code is applied. By this process, the energy minimum is determined evolving from a chosen initial magnetization state, either a SD or a PVC state. Then, the MGS determined by this energy minimization process is either a SD-O mixed state or a PVC state. The O-state and the SD state are then further distinguished from the SD-O mixed state by another criterion, which is an analytical expression discussed in detail in section III. In the above procedure to determine the MGS numerically with different particle size, the particle radius is varied by the step size of 1 to 2 nm. It leads to a possible uncertainty of 0.5 to 1 nm to determine the phase boundary position in the phase diagram.

## 3. Onion state by analytical expression and numerical analysis

In this section, the equations for the magnetization configuration and the corresponding total energy of a 3D O-state are proposed and studied in detail. We first select the spherical coordinate system, $\vec{r}(r, \theta, \phi)$, with the origin located at the center of the hollow sphere, in which, $r$ is the radial, $\theta$ is the zenith and $\phi$ is the azimuthal coordinates. The magnetization is treated as a continuous vector function $\vec{M}(\vec{r})$, so that $\vec{M}(\vec{r})\delta V$ gives the total magnetic moment within the volume $\delta V$ centered at the position $\vec{r}$, as shown by the simple diagram in Fig. 2. The magnetization vector is then expressed as $\vec{M}(\vec{r}) = M_r(\theta)\hat{r} + M_\theta(\theta)\hat{\theta}$. Obviously, $\vec{M}(\vec{r})$ does not depend on the azimuthal angle $\phi$ owing to the symmetric structure of the O-state with respect to the polar axis. In addition, the possible dependence of the magnetization distribution on the radial coordinate $r$ is neglected. The magnitude of error arising from this approximation in the determination of the total energy for an O-state is small. It will be assessed in Sec. IV. To simplified the treatment, the magnetization is normalized to the saturation value, $M_S$, and expressed as $\vec{m}(\vec{r}) = \vec{M}(\vec{r})/M_S$, $m_r(\vec{r}) = M_r(\vec{r})/M_S$, and $m_\theta(\vec{r}) = M_\theta(\vec{r})/M_S$. The magnetization configuration of a 3D O-state viewed in the 2D cross sectional plane, which contains the polar axis as shown in Fig. 2, is the same as that of a 2D nanoring. By adapting the treatment for the 2D nanoring by Landeros et al.,[21] the normalized magnetization components in half of the 2D circular ring can be written as,

$$m_r(\alpha,\theta) = \begin{cases} f(\alpha,\theta), & 0 < \theta < \pi/2 \\ -f(\alpha,\pi-\theta), & \pi/2 < \theta < \pi \end{cases}, \tag{1a}$$

$$m_\theta(\alpha,\theta) = \begin{cases} -\sqrt{1-f^2(\alpha,\theta)}, & 0 < \theta < \pi/2 \\ -\sqrt{1-f^2(\alpha,\pi-\theta)}, & \pi/2 < \theta < \pi \end{cases}, \tag{1b}$$

where $f(\alpha,\theta) = \cos^\alpha(\theta)$ with $\alpha \geq 1$. With the half ring rotating around the polar axis for an angle of $360^o$, the 3D O-state is then obtained.

Equations (1a) and (1b) actually describe the magnetization configuration for a SD-O mixed state of the hollow sphere. With $\alpha = 1$, the magnetization is uniformly distributed and directed in parallel to the polar axis, forming a SD state. With $\alpha > 1$, however, the direction of magnetization at different local points exhibits different degrees of deviation away from the direction of polar axis, forming an O-state. In the large limit with $\alpha \gg 1$, $m_r(\alpha, \theta)$ approaches zero except in the regions around the two poles. Consequently, the magnetization completely follows the curvature of the shell except in the vicinity of the two poles, which form a magnetization core structure similar to that with the VC state. According to the above analysis, an O-state has a distorted magnetization configuration from the SD state. Therefore, it is practical to define a criterion according to the analytical expressions, Eqs. (1a) and (1b), that an O-state is with $\alpha > 1$ while a SD state is with $\alpha = 1$.

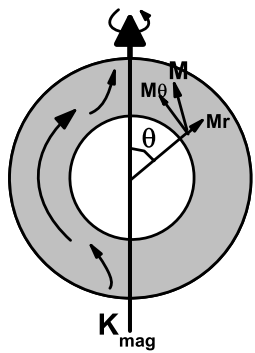

FIG. 2 Simple diagram for an onion state.

By Eqs. (1a) and (1b), the total energy, $E_O(\alpha)$, of a SD-O state as a function of $\alpha$ parameter is expressed as follows:

$$E_O = E_{ex} + E_m + E_K, \tag{2a}$$

$$E_{ex} = A \int_V [(\nabla m_x)^2 + (\nabla m_y)^2 + (\nabla m_z)^2] dV, \tag{2b}$$

$$E_m = \frac{\mu_0 M_S}{2} \int_V \vec{m}(\vec{r}) \cdot \nabla U(\vec{r}) dV, \tag{2c}$$

in which $U(\vec{r}) = -\frac{M_S}{4\pi} \int_V \frac{\nabla' m(\vec{r}')}{|\vec{r} - \vec{r}'|} dV' + \frac{M_S}{4\pi} \int_S \frac{\hat{n}' \cdot m(\vec{r}')}{|\vec{r} - \vec{r}'|} dS'$ is the magnetostatic potential, and

$$E_K = K_{mag} \int_V \{1 - [m(\vec{r}) \cdot \hat{n}_{mag}]^2\} dV. \tag{2d}$$

In Eqs. (2a)-(2d), $E_{ex}$ is the exchange energy with the exchange constant $A$, $E_m$ is the magnetostatic energy, and $E_K$ is the anisotropy energy with the direction of anisotropy axis defined by $\hat{n}_{mag}$. To determine the MGSs from the SD-O mixed state, it is necessary to find out explicitly the state of lowest total energy by $E_O(\alpha)$. Figure 3 shows two typical calculation results for the energy density, i.e., $E_O(\alpha)$ divided by the shell volume of the hollow sphere, $V_{sphere}$. For a smaller nanosphere with $R = 10$ nm, $\Lambda_R = 0.2$ and $K_{mag} = 4.0 \times 10^5$ J/m$^3$, there is no minimum point in $E_O(\alpha)$. The smallest value of $E_O(\alpha)$ occurs at $a = 1$, as shown by the curve with solid circles in Fig. 3. It indicates that the MGS is a SD state. On the other hand, for the nanosphere with $R = 29$ nm, $\Lambda_R = 0.4$ and $K_{mag} = 4.0 \times 10^5$ J/m$^3$, a local minimum appears at $\alpha = 1.2$. It indicates that the MGS is an O-state.

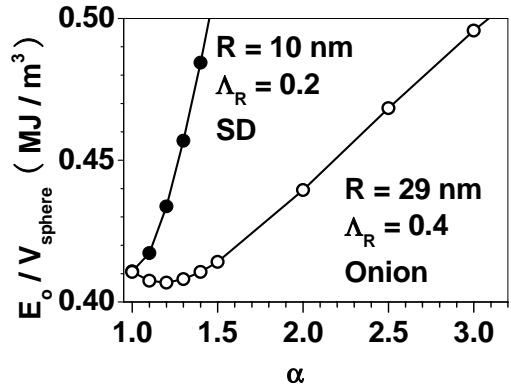

FIG. 3 Energy density for the SD and the O states versus the parameter, $\alpha$.

To further confirm the validity of the total energy calculated for the SD or O-state by Eqs. (2a)-(2d) and to determine the phase boundary line separating these two states, the total energy has been calculated also by another independent pathway, i.e., by direct calculations using the micromagnetic simulation program without resorting to the analytical expressions of Eqs. (1a) and (1b). Figure 4(a) shows the total energy versus $R$ for the Co hollow sphere with $K_{mag} = 4.0 \times 10^5$ J/m$^3$. Two different values of $\Lambda_R$, i.e., 0.4 and 0.8, are calculated by both the analytical, $E_A$, and numerical, $E_N$, approaches. The results obtained by these two approaches are almost the same. The ratio, $E_A/E_N$, is plotted in the inset of Fig. 4(a). It is almost equal to 1, with a maximum difference of 1%, for $\Lambda_R = 0.4$. In the thin-shelled limit with $\Lambda_R = 0.8$, the difference becomes slightly larger, with ($E_A/E_N$ - 1) ranging from 3 to 7%. Similar result has also been obtained for the Co hollow sphere with $K_{mag} = 0$, as shown in Fig. 4(b). The result calculated by the analytical expression in both cases always gives a slightly higher energy value. This is mainly attributed to the approximation that the possible dependence of the magnetization distribution on the radial coordinate $r$ is neglected in Eqs. (1a) and (1b). The difference becomes more pronounced with the thin-shelled particle. However, even in the extreme condition, the difference remains a higher order effect, less than 7% in Fig. 4(a). This indicates that the analytical expressions by Eqs. (1a) and (1b) are valid for the onion state under the present study.

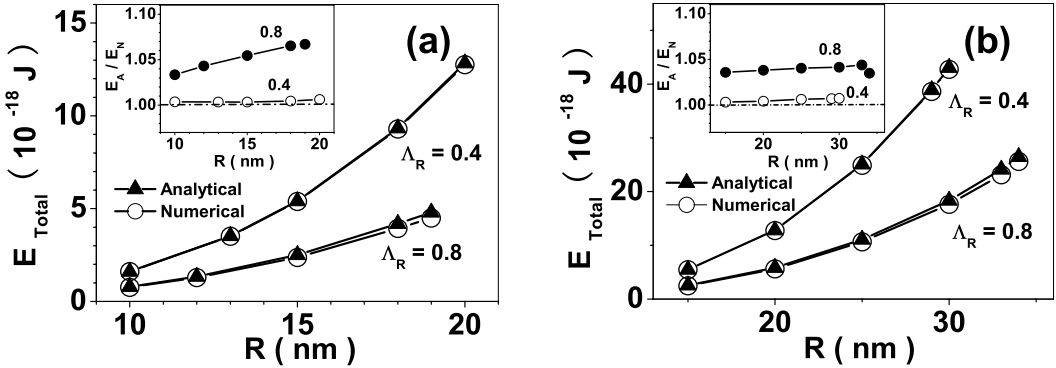

FIG. 4 Total energy calculated for the onion and the SD states by two approaches, one according to Eqs. (1a) and (1b), as shown by the solid triangles, and the other by a direct micromagnetic simulation method, as shown by the open circles. (a) For hollow sphere with $K_{\mathrm{mag}} = 4.0 \times 10^5$ J/m$^3$, and (b) with $K_{\mathrm{mag}} = 0$.

### 4. Phase diagrams

The phase diagrams are determined by the numerical process of energy minimization in addition to the criterion discerning the O-state from the SD state. For the single crystal Co nanosphere, the phase diagram is shown in Fig. 5. By the iteration process of energy minimization, it is found that both the initial PVC and SD states would eventually evolve into a SD-O mixed state if $R$ is smaller than a critical radius, $R^{cri}$, which is around 15 nm or so, as shown in Fig. 5. It indicates that for a small particle of single crystal with $R < R^{cri}$, no energy barrier exists to prevent the transition from a PVC state into a SD-O mixed state. However, with the particle radius exceeding $R^{cri}$, an energy barrier is present. It separates the two local minima in energy corresponding to the SD-O mixed state and the PVC state. In order to determine the MGS and the corresponding phase boundary for a particle with $R > R^{cri}$, the values of these two local minima are calculated explicitly for comparison. The phase line separating the PVC from the SD-O mixed state thus determined is represented by the curve with open triangles in Fig. 5. The locations of the symbols in the diagram are for the calculation points while the phase boundary line is obtained by interconnecting the adjacent symbols. Therefore, metastable states are present at a finite temperature in the region of $R > R^{cri}$. The other phase line, which separates the SD from the O-state, is represented by the curve with the open circles. According to this phase diagram, the SD state is in favor of a small particle with a thick shell with which the exchange energy dominates, whereas the O-state prevails over the SD state for a small particle with a thin shell. On the other hand, for a large particle, the magnetization is in favor of the PVC state with a magnetization flux closure structure to reduce the magnetostatic energy.

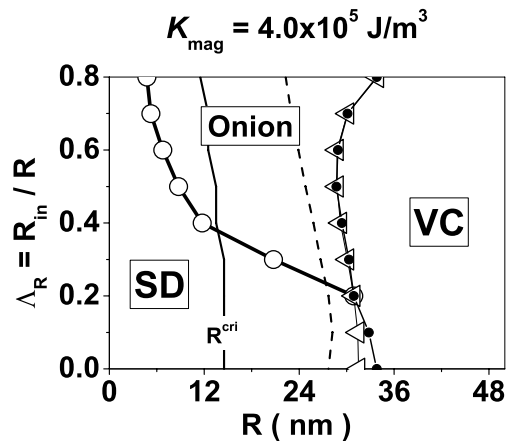

FIG. 5 Phase diagram for a nanosized Co hollow sphere with $K_{mag} = 4.0 \times 10^5$ J/m$^3$. The phase line with open triangles separates the PVC and the SD–O mixed states and the phase line with open circles separates the SD and O-states. The line

with solid circles is calculated for the AVC state, which has a higher energy than the PVC state. The short dash curve below the one with open triangles is for the result by the analytical approach (Ref. 11). The solid curve below the dash one is for $R = R^{\text{cri}}$.

The phase line separating the simple VC state from the SD state was investigated analytically by Goll et al[11] as represented by the dash line in Fig. 5. For the simple VC state, the magnetic coupling energy between the two vortex cores is not taken into consideration. In the present work, however, both of the PVC and AVC states are calculated. For $\Lambda_R = 0$, i.e. for a solid Co nanoparticle, the phase boundary separating the SD state from the PVC state is $R = 34.0$ nm according to our calculation. It is larger than the value of 27.7 nm obtained by the analytical work by Goll et al,[11] but agrees well with the calculated value of 34.1 nm by Aharoni.[25] The curve with the solid circles plotted in Fig. 5 is for the phase boundary if the energy comparison is made between AVC and the SD-O mixed state. Both phase lines coincide at $\Lambda_R > 0.2$ and separate only at $\Lambda_R < 0.2$. The difference in the total energy between these two states is small, calculated as 8% according to $(E_{AVC} - E_{PVC})/0.5(E_{AVC} + E_{PVC})$ for $\Lambda_R = 0$.

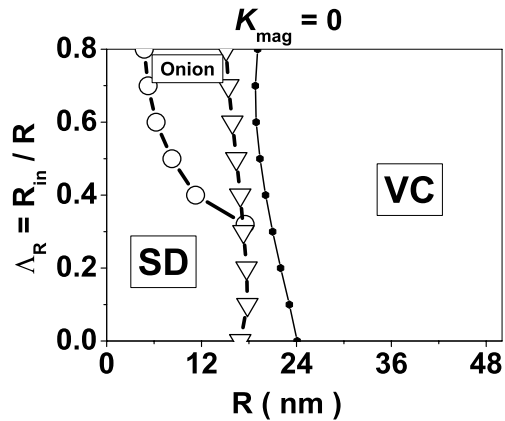

FIG. 6 Phase diagram for a nanosized Co hollow sphere with $K_{\text{mag}} = 0$. The phase line with open triangles separates the PVC and the SD–O mixed states and the phase line with open circles separates the SD and O-states. The line with solid circles is calculated for the AVC state.

For the nanosphere with $K_{mag} = 0$, the phase diagram is plotted in Fig. 6. The position of the phase line separating the PVC state from the SD-O mixed state is at $R$ ~18 nm, much smaller than 34.0 nm shown in Fig. 5 for a single crystal Co hollow sphere. In addition, it coincides with the line, $R = R^{cri}$, owing to the absence of the anisotropy energy. It indicates that there is no metastable state for the PVC and the SD-O mixed states as discussed for the nanosphere with $K_{mag} > 0$. In the region for the SD-O mixed state, O-state shows up as the shell thickness decreases from the SD region, similar to the behavior with the single crystal hollow sphere. For the AVC state, the phase line shifts significantly to the large particle end from that for the PVC state. It reflects that the energetic difference between the AVC and PVC states due to the interaction between the two vortex cores becomes pronounced with the absence of the anisotropy energy. In addition, there is an energy barrier between these two states, which is attributed to the reversion of one of the two vortex cores from the antiparallel to the parallel configuration. This makes the AVC state a metastable one.

According to the present calculation, the energy difference between the PVC and the AVC states is comparatively small for a single crystal Co hollow sphere in the presence of the magnetic anisotropy. The distinction between the boundary lines determined for the PVC and AVC states shows a maximum difference of 2.3 nm for a solid particle with $\Lambda_R = 0$, as shown in Fig. 5. This difference is much smaller than the variation of the phase boundary line, which separates the VC and SD-O states, from $R = 34$ nm ($\Lambda_R = 0$) to $R = 18$ nm attributed to the absence of the magnetic anisotropy. It justifies the assumption in the previously reported analytical investigation[11] that the interaction between the two vortex cores is negligible. However, the above assumption is not valid for a polycrystalline sample with $K_{mag} = 0$. The phase line for the AVC state moves significantly toward the large particle side from that for the PVC state, as shown in Fig. 6.

## 5. Discussion

The characteristic length scales that are important to describe the equilibrium magnetic properties of nanomagnetic material include the exchange length, $L_{ex}$, and the critical SD radius, $R_{SD}$. The exchange length, $L_{ex}$, is the length scale below which the atomic exchange coupling strength dominates the magnetostatic interaction, while the single domain size, $D_{SD} = 2R_{SD}$, reflects the competition between the magnetostatic energy and the formation energy of domain wall.[26,27] For a spherical hcp Co particle, it has been reported that $L_{ex}^{Co}$ ~ 5 nm,[13] and $R_{SD}$ ~ 34 nm.[25] In the present work, the critical single domain radius for a Co nanosphere of single crystal is $R_{SD}$ ~ 34.0 nm. It is the same as the previously reported value.[25] However, for the polycrystalline nanosphere, assuming $K_{mag} = 0$, $R_{SD}$ is obtained numerically as 18.0 nm without accounting for the anisotropy energy expressed by Eq. (2d). For a polycrystalline nanosphere, in general, the value of the anisotropy constant is expected to fall between the bulk value and 0. Therefore, $R_{SD}$ for a polycrystalline nanosphere of Co is expected to range from 18 to 34 nm according to the present calculation.

For a dynamical process of magnetization reversal in response to a sweeping field, the relevant magnetic length scale is the coherence length, $L_{coh}$. Usually, the coherence length is larger than the exchange length, $L_{ex}$, by a factor depending on the geometry and material of the particle under consideration, and is smaller than the SD size, $L_{SD} = 2R_{SD}$.[26,27] In particular, for a hard ferromagnet, $L_{SD}$ could be larger than $L_{coh}$ by more than 1 order of magnitude. The coherence length for a solid spherical particle of Co is $5.099\sqrt{A/\mu_0 M_S^2} = 17.8$ nm,[27] and determined as 16.4 nm in one of the recent experiments.[9] This value is smaller than $L_{SD} = 2R_{SD} = 36$ nm for a polycrystalline sample and is even much smaller than $L_{SD} = 64$ nm for a particle of single crystal. For a Stoner-Wohlfarth (SW) type Co nanosphere, *i.e.* a Co nanosphere with a radius $R < L_{coh}/2$, roughly equal to 8 or 9 nm, the magnetization reversal mode is by the coherence rotation. At a finite temperature in response to a sweeping field, the blocking temperature, $T_B$, for a hcp Co nanosphere with $K_{mag} = 4.0 \times 10^5$ J/m$^3$, is determined according to the expression, $K_{mag}V \approx 25 k_B T_B$. Within the temperature range, $T_B < T < T_C$, the SW nanosphere would exhibit superparamagnetic (SPM) property. The blocking temperature for a hollow sphere is expected to reduce from the value of the same-sized solid particle by the factor, $1 - (\Lambda_R)^3$, due to the reduction in the particle volume. For a thin-shelled SW particle of radius $R < 7$ nm with $\Lambda_R = 0.8$, for instance, its volume is smaller than that of a solid particle with the same radius by a factor of 0.488. The blocking temperature is expected to be lower by the same factor accordingly. On the other hand, $K_{mag}$ is assumed as 0 in the calculation for the phase diagram of polycrystalline Co nanosphere. This implies that the blocking temperature is $T_B = 0$, and the polycrystalline Co nanoparticles would exhibit SPM property at any finite temperature, $T > 0$ K.

With the particle size approaching the small limit, a great enhancement of the magnetic anisotropy has been widely reported.[28,29] It is attributed to the reduction of the crystal symmetry and the surface anisotropy effects, and is sample dependent.[29] For a Co nanocluster formed by 30 atoms, which is about 0.5 nm in radius, the magnetic anisotropy reaches as high as $3.0 \times 10^7$ J/m$^3$.[29] It is higher than the bulk value by two orders. Although the enormous enhancement of magnetic anisotropy with a small particle is sample dependent, a number of significant results are expected. First, for a particle with $R < 1$ nm or so, the blocking temperature is expected to be much higher than the value predicted by the expression $K_{mag}V \approx 25 k_B T_B$ using the bulk magnetocrystalline anisotropy, $K_{mag}$. This would further reduce the temperature interval between $T_B$ and $T_C$ for the observation of SPM property, in addition to the reduction of $T_C$ attributed to the finite size effect as discussed in Ref 30. Second, the phase diagrams presented in the present work is expected to change substantially for the particle with the shell thickness less than 1 nm since the surface anisotropy energy is not accounted for in the calculation of the total energy by Eq. (2d). By the above discussion, the phase boundary lines in the phase diagrams are expected to deviate from the present result in the thin-shelled and small-sized limits.

According to the present analysis, there is a potential barrier between the PVC and SD-O states for the particle of $K_{mag} > 0$ with $R > R^{cri}$. This results in the presence of a metastable state not shown in the phase diagram by Fig. 5. Similar stability problem associated specifically with the VC and SD states arising from a potential barrier between these two states has been investigated for 2D nanodots of Co at a finite temperature by a previous experiment.[31] For the 3D hollow sphere with $K_{\text{mag}} > 0$ and $R > R^{\text{cri}}$, we have estimated the energy barrier separating the PVC and the SD-O mixed states. For nanosphere of $\Lambda_R = 0.5$ and $R = 22$ nm, the energy barrier is about 680 K, and for a nanosphere of $\Lambda_R = 0.5$ and $R = 24$ nm, it is about 2000 K. The latter is even higher than the Curie temperature of Co, $T_C = 1388$ K. It indicates that the thermal activation effect alone is not enough to cause a transition across the potential barrier from the metastable state to reach the real MGS.

## 6. Conclusion

An onion state is proposed and confirmed as a MGS for a 3D hollow nanoparticle of Co, in addition to the previously known SD and VC states. Meanwhile, an analytical equation is obtained for the magnetization configuration of the 3D onion state. The total energy of the onion state calculated according to the proposed analytical equation is shown to be consistent with the result by the direct numerical calculation. Two phase diagrams are determined for the SD, VC, and the newly introduced onion states. One of the phase diagrams is for the single crystal Co nanosphere with $K_{mag} = 4.0 \times 10^5$ J/m$^3$, and the other is for the polycrystalline Co nanosphere with $K_{mag} = 0$. Significant difference is revealed between the two diagrams owing to the presence of the uniaxial anisotropy. In particular, the critical SD radius for the nanosphere with $K_{mag} = 4.0 \times 10^5$ J/m$^3$ is almost twice the value for the one with $K_{mag} = 0$.

Acknowledgements

This work is supported in part by the National Fund for Fostering Talents of Basic Science (NFFTBS-J0630311).